\documentclass[prb,twocolumn,superscriptaddress,amscd,amssymb,verbatim]{revtex4-1}

\usepackage{graphicx}
\usepackage{textcomp}
\usepackage[OT1,T1]{fontenc}
\usepackage{color}
\definecolor{red}{rgb}{1,0,0}

\usepackage{epstopdf}

\newcommand{\Cu}{CuMnO$_2$}
\newcommand{\Na}{$\alpha$-NaMnO$_2$}

\begin{document}

\title{Magnetic inhomogeneity on a triangular lattice:\\
the magnetic-exchange versus the elastic energy and the role of disorder}
\author{A. Zorko}
\email{Correspondence and request for materials should be addressed to A. Z. (andrej.zorko@ijs.si)}
\affiliation{Jo\v{z}ef Stefan Institute, Jamova c.~39, 1000 Ljubljana, Slovenia}
\author{J. Kokalj}
\affiliation{Jo\v{z}ef Stefan Institute, Jamova c.~39, 1000 Ljubljana, Slovenia}
\author{M. Komelj}
\affiliation{Jo\v{z}ef Stefan Institute, Jamova c.~39, 1000 Ljubljana, Slovenia}
\author{O. Adamopoulos}
\affiliation{Institute of Electronic Structure and Laser, Foundation for Research and Technology -- Hellas, Vassilika Vouton, 71110 Heraklion, Greece}
\author{H. Luetkens}
\affiliation{Laboratory for Muon Spin Spectroscopy, Paul Scherrer Institute, CH-5232 Villigen, Switzerland}
\author{D. Ar\v{c}on}
\affiliation{Jo\v{z}ef Stefan Institute, Jamova c.~39, 1000 Ljubljana, Slovenia}
\affiliation{Faculty of Mathematics and Physics, University of Ljubljana, Jadranska c.~19,
1000 Ljubljana, Slovenia}
\author{A. Lappas}
\affiliation{Institute of Electronic Structure and Laser, Foundation
for Research and Technology -- Hellas, Vassilika Vouton, 71110 Heraklion, Greece}

\date{\today}
\begin{abstract}
Inhomogeneity in the ground state is an intriguing, emergent phenomenon in magnetism. 
Recently, it has been observed in the magnetostructural channel of the geometrically frustrated $\alpha$-NaMnO$_2$, for the first time in the absence of active charge degrees of freedom. 
Here we report an in-depth numerical and local-probe experimental study of the isostructural sister compound CuMnO$_2$ that emphasizes and provides an explanation for the crucial differences between the two systems.
The experimentally verified, much more homogeneous, ground state of the stoichiometric CuMnO$_2$ is attributed to the reduced magnetoelastic competition between the counteracting magnetic-exchange and elastic-energy contributions.
The comparison of the two systems additionally highlights the role of disorder and allows the understanding of the puzzling phenomenon of phase separation in uniform antiferromagnets.
\end{abstract}
\maketitle

Although phase separation in a uniform system is a widespread phenomenon in diverse fields of matter
\cite{seul_domain_1995, malescio_stripe_2003, dagotto_complexity_2005}, ranging from biological systems  \cite{seifert_configurations_1997, baumgart_imaging_2003, kondo_reaction-diffusion_2010},
 to soft matter \cite{maclennan_novel_1992,harrison_mechanisms_2000}, and strongly correlated electron systems \cite{tranquada_evidence_1995, dagotto_colossal_2001, roger_patterning_2007, vojta_lattice_2009, park_electronic_2009, bauer_electronic_2011, seo_disorder_2013},
 in magnetism the microscopic pattering has been, until recently, almost exclusively limited to thin ferromagnetic (FM) films \cite{debell_dipolar_2000, portmann_inverse_2003}.
In this case, such a pattering is a trade-off between minimizing the exchange and the dipolar energies. It thus represents one possible manifestation of a general requirement of multiple competing phases that can lead to inhomogeneous states.
Lately, it has become increasingly apparent that a similar competition between energetically nearly equivalent phases is also responsible for phase separation in geometrically frustrated spin systems \cite{zorko_frustration-induced_2014, de_groot_competing_2012, kamiya_formation_2012, nakajima_microscopic_2012} that are generically characterized by ground-state degeneracy \cite{lacroix_introduction_2011}.
However, the balance between the competing phases in these systems is generally much more delicate and, therefore, poorly understood.

Recently, the spatially anisotropic triangular antiferromagnet \Na,~with dominant intrachain ($J_1$) and geometrically frustrated interchain ($J_2$) antiferromagnetic (AFM) exchange interactions (inset in Figure~\ref{fig3}a), has been highlighted as a paradigm of a phase-separated ground state in the absence of active charge degrees of freedom \cite{zorko_frustration-induced_2014}. 
Its AFM order that sets in below the N\'eel temperature $T_{\rm N}=45$~K is accompanied by a simultaneous structural deformation \cite{giot_magnetoelastic_2007}.
This was initially suggested as being a phase transition from the high-temperature monoclinic ($C2/m$) to the low-temperature triclinic ($P\bar{1}$) crystal structure \cite{giot_magnetoelastic_2007}.
However, more detailed, recent experiments have shown that the magnetic order fails to drive this improper ferroelastic transition to completion \cite{zorko_frustration-induced_2014}.
Instead, an intricate magnetostructurally inhomogeneous state on the nano-scale has been discovered below the $T_{\rm N}$. 
Such a state was suggested to be an unforeseen consequence of the subtle interplay between the geometrical frustration and the competing structural phases \cite{zorko_frustration-induced_2014}.

In order to fully understand this novel phenomenon, further theoretical studies and experimental investigations  of related compounds are of paramount importance. 
In this respect, a comparison with the crystallographically \cite{Kondrashev_1959} and magnetically \cite{doumerc_magnetic_1994} analogous sister compound \Cu, known as the mineral crednerite, is particularly relevant.
Here, in contrast to \Na, the emergent magnetic order below $T_{\rm N}=65$~K is believed to lift the macroscopic degeneracy in the spin space completely, by inducing the monoclinic-to-triclinic structural phase transition \cite{damay_spin-lattice_2009}. 
This spin-induced phase transition is witnessed by the splitting of several families of nuclear Bragg reflections \cite{vecchini_magnetoelastic_2010,zorko_frustration-induced_2014}. 
It was suggested to reflect the strong magnetoelastic (ME) coupling that allows for the development of shear strain at a low energy cost \cite{vecchini_magnetoelastic_2010}.
Interestingly, the strain is significantly enhanced \cite{poienar_substitution_2011} in the off-stoichiometric \cite{trari_preparation_2005} Cu$_{1+x}$Mn$_{1-x}$O$_2$, where  $T_{\rm N}$ is reduced and the structural transition temperature is further suppressed with increasing $x$ even for small doping levels  \cite{poienar_substitution_2011,garlea_tuning_2011,terada_magnetic_2011}.
Moreover, in off-stoichiometric samples the interlayer ordering changes from AFM, observed in stoichiometric \Cu, to FM \cite{poienar_substitution_2011, garlea_tuning_2011, terada_magnetic_2011}, which was attributed to a partial substitution of Cu$^{2+}$ for Mn$^{3+}$ that should effectively change the interlayer exchange coupling from AFM to FM\cite{ushakov_orbital_2014}.
On the other hand, this implies that \Cu~may be very close to an electronic instability, possibly of a similar kind to that found in stoichiometric \Na. 

Obtaining in-depth information about the magnetic and structural properties of \Cu~on the local scale should clarify the differences with respect to the isostructural \Na.
Such knowledge would also help to address the pending issue of the microscopic origin of the phase-separation phenomenon in geometrically frustrated magnets.
The most obvious ambiguities arise from questions like: why is the $T_{\rm N}$ enhanced in \Cu~compared to \Na, despite the theoretically predicted sizably smaller exchange interactions in the former compound \cite{jia_magnetic_2011}; what is the role of the ME coupling in establishing the structural distortion below the $T_{\rm N}$, what is the role of disorder; and ultimately, why does the structural phase transition appear to be fully developed in \Cu, while in \Na~it only manifests in a phase-separated state. 
Here, we answer these questions by combining numerical calculations with local-probe experimental investigations. 
First, we determine the dominant intralayer exchange interactions by modelling the magnetic susceptibility via exact-diagonalization calculations.
We demonstrate that the difference in the $T_{\rm N}$ for the two compounds can be understood via the mismatch of the two non-equivalent interchain exchange interactions, i.e., by the different extent of the frustration present in the two compounds, while the ME contribution is negligible in this respect.
Moreover, we provide the first experimental microscopic insight into the magnetism of \Cu~via $^{63, 65}$Cu nuclear magnetic resonance (NMR) and nuclear quadrupolar resonance (NQR) measurements, as well as complementary muon spin relaxation ($\mu$SR) measurements.
These experiments clearly reveal that the ground state is more homogeneous in the Cu case than in the Na case and suggest that the stoichiometric \Cu~is on the verge of a phase-separation instability.   
Finally, our {\it ab-initio} calculations suggest that the more homogeneous state of the stoichiometric \Cu~originates from an enhanced energy difference (when compared to \Na) between the two competing phases, born out of the magnetic-exchange and the elastic-energy changes below the $T_{\rm N}$.

\section*{Results}
\begin{figure}[t]
\includegraphics[trim = 0mm 0mm 0mm 0mm, clip, width=1\linewidth]{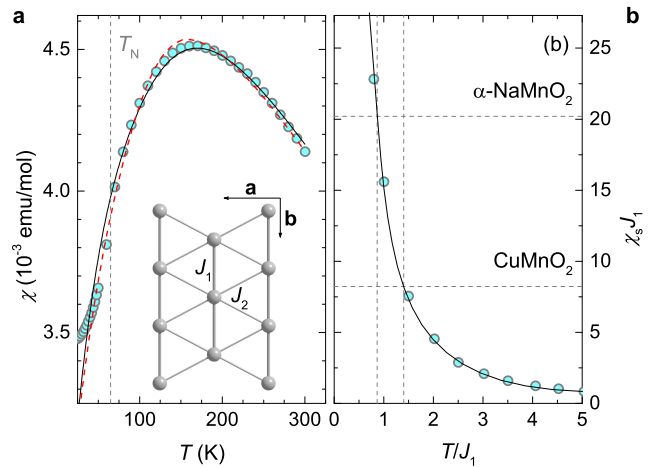}
\caption{{\bf Determination of $J$'s and $T_{\rm N}$.} {\bf (a)} The magnetic susceptibility $\chi=M/H$ ($M$ is the magnetization and $H$ is the applied magnetic field) of \Cu, measured by a SQUID magnetometer in a field of $\mu_0 H=0.1$~T. 
The solid and dashed lines denote the best FTLM fits with the average and the scaled curves, respectively (see Methods for details). 
The former yields the exchange-coupling constants $J_1=53.5$~K, $J_2/J_1=0.25$, and the latter $J_1=52.1$~K, $J_2/J_1=0.29$. 
Inset shows the spatially anisotropic triangular spin lattice of the \Cu~in the monoclinic setting, with intrachain $J_1$ (thick bonds) and interchain $J_2$ (thin bonds) exchange constants. 
{\bf (b)} The temperature dependence of the staggered susceptibility $\chi_{\rm s}$ multiplied by $J_1$ for spin-2 chains (adopted from Ref.~\onlinecite{kim_monte-carlo_1998}). The solid line is a guide to the eye. 
The N\'eel transition temperatures $T_{\rm N}=1.40 J_1=74$~K in \Cu~and $T_{\rm N}=0.87 J_1=57$~K in \Na~are predicted (dashed lines) by equation~(\ref{eq5}).}
\label{fig3}
\end{figure}

\noindent {\bf Determination of the dominant exchange interactions and the $T_{\rm N}$.}
In order to understand the apparently significantly different properties of the two isostructural compounds, \Cu~and \Na, a proper determination of the dominant terms in the Hamiltonian is crucial.
Therefore, we applied numerical finite-temperature Lanczos method (FTLM) simulations and density-functional theory (DFT) calculations.
The former were aimed at quantifying the two main magnetic exchange interactions ($J_1$ and $J_2$) of the isotropic Heisenberg model on the spatially anisotropic two-dimensional (2D) triangular lattice (inset in Figure~\ref{fig3}a),
\begin{equation}
\mathcal{H}=J_1\sum_{(ij)}{\bf S}_i \cdot {\bf S}_j+J_2\sum_{[kl]}{\bf S}_k \cdot {\bf S}_l.
\label{eq4}
\end{equation}
Here, the first sum runs over the (stronger) intrachain bonds in one direction, while the second sum runs over the (weaker) interchain bonds in the other two directions on the triangular lattice of spins $S=2$. 

The temperature-dependent magnetic susceptibility $\chi(T)$ of this model was calculated for various $m \times n$ spin clusters (see Methods).
However, even for the largest reachable cluster sizes ($7\times 2$), some finite-size effects remain present at low $T$.
With the use of results from many clusters ($m=2-7$) these effects can, however, be reduced and the value of $\chi$ at the thermodynamic limit is thus approached. 
Both, the average susceptibility curve from the two largest-size clusters (that gives a better approximation than the individual clusters; see Methods) and the susceptibility curve obtained in an approach similar to finite-size scaling (see Methods) fit the experimental data above the $T_{\rm N}$ very well (Figure~\ref{fig3}a). 
A disagreement with the experimental data below the $T_{\rm N}$, on the other hand, is expected, because the 2D Heisenberg model cannot account for a finite $T_{\rm N}$.
Both approaches yield very similar exchange-coupling constants, which we estimate to be $J_1=53$~K and $J_2/J_1=0.27(2)$.
These are in very good agreement with recent {\it ab-initio} predictions \cite{jia_magnetic_2011}, $J_1=56$~K and $\bar{J}_2/J_1=0.23$, which, in principle, could be erroneous due to the unknown on-site repulsion \cite{ushakov_orbital_2014}. 

Our calculations thus confirm that the exchange interactions are indeed reduced for \Cu~compared to \Na, where \cite{zorko_magnetic_2008} $J_1=65$~K and $J_2/J_1=0.44$. 
Despite this fact, in \Cu~the $T_{\rm N}$ is increased with respect to that found in \Na~by more than 40\%. 
So far, this has been attributed to a difference in the interlayer coupling \cite{damay_spin-lattice_2009} $J'$, which, however, is rather small \cite{jia_magnetic_2011,ushakov_orbital_2014} and can, therefore, only slightly affect the $T_{\rm N}$ on 2D Heisenberg lattices \cite{yasuda_netemperature_2005}.
Furthermore, the amount of frustration reflected in the $J_2/J_1$ ratio should directly influence the $T_{\rm N}$, as the frustration is known to suppress the spin correlations \cite{zheng_temperature_2005}.
In \Na~and \Cu~the intrachain exchange coupling is dominant. 
Therefore, these compounds can be regarded as systems of coupled spin chains, which is manifested in the one-dimensional character of the magnetic excitations in $\alpha$-NaMnO$_2$ \cite{stock_one-dimensional_2009}. 
For such systems the $T_{\rm N}$ can be determined with the use of a random-phase approximation \cite{scalapino_1975, schulz_dynamics_1996}. 
In this approach the interchain coupling is treated at the mean-field level, whereas the intrachain interactions are treated exactly. 
For isotropic interchain coupling in a non-frustrated lattice the $T_{\rm N}$ is determined by the condition\cite{yasuda_netemperature_2005,schulz_dynamics_1996,scalapino_1975,irkhin_calculation_2000}
$zJ_2\chi_s(T_{\rm N})=1$,
where $\chi_s(T)$ is the chain's staggered susceptibility. 
Within this approach, we generalize the above condition for $T_{\rm N}$ to include the two interchain constants ($J_{\rm 2a}>J_{\rm 2b}$) pertinent to the triclinic phase of \Cu~and \Na, as well as the interlayer coupling $J'$;
\begin{equation}
\chi_{\rm s}(T_{\rm N})=\frac{1}{k[z\left(J_{\rm 2a}-J_{\rm 2b} \right)+z'J']}.
\label{eq5}
\end{equation}  
Here, $z=z'=2$ corresponds to the number of neighbouring coupled chains and planes, respectively, while $J_{\rm 2b}$ adopts the minus sign because it frustrates the AFM order dictated by the larger $J_{\rm 2a}$.
The constant $k$ renormalizes the coordination numbers and is reduced from unity \cite{irkhin_calculation_2000, bocquet_finite-temperature_2002} because of quantum effects \cite{yasuda_netemperature_2005}. 
As $\left( J_{\rm 2a}-J_{\rm 2b}\right)/J_1 =0.083$ and 0.035 in \Cu~and \Na, respectively \cite{jia_magnetic_2011}, while $J'/J_1$ is expected to be an order of magnitude smaller \cite{jia_magnetic_2011,ushakov_orbital_2014}, 
we estimate the $T_{\rm N}$ by neglecting $J'$ in equation~(\ref{eq5}) and by taking $k=0.7$, which is appropriate for quasi-one dimensional cases (see Fig.~2 in Ref.~\onlinecite{yasuda_netemperature_2005}).
This gives $T_{\rm N}=74$~K in \Cu~and $T_{\rm N}=57$~K in \Na~(see Figure~\ref{fig3}b), which are in good agreement with the experimental values of 65~K and 45~K, respectively. 
We note that for anisotropic interchain couplings the constant $k$ is expected to be further reduced and, ultimately, for \cite{irkhin_calculation_2000, bocquet_finite-temperature_2002} $J'\to 0$ also $k\to 0$, leading to $T_{\rm N}\to 0$, which is consistent with the Mermin-Wagner theorem (no long-range order in the 2D Heisenberg model at finite $T$). 
However, it has been shown \cite{Siurakshina_l._theory_2000} that for quasi-2D systems the dependence of the $T_{\rm N}$ and, in turn also of $k$, on $J'$ is sub-logarithmic. 
Therefore, for the exchange-coupling constants related to \Cu~and \Na, $k$ will be somewhat, but not drastically, reduced from the value $0.7$, which is perfectly in line with the small theoretical overestimates of the $T_{\rm N}$.
 
This analysis reveals that the N\'eel transition is predominantly determined by the Heisenberg Hamiltonian of equation~(\ref{eq4}). 
Moreover, the ferrodistortive structural transition accompanying the magnetic ordering and leading to the splitting of the two interchain exchange constants ($J_{2a}$, $J_{2b}$) in the triclinic phase is needed to ensure a finite $T_{\rm N}$. 
The extent to which frustration is relieved in the triclinic phase of the \Cu~elevates its ordering temperature above the ordering temperature in \Na. 
Other factors, such as the interlayer coupling, the magnetic anisotropy and the ME coupling, can, at best, only slightly shift the $T_{\rm N}$. 
\begin{figure}[t]
\includegraphics[trim = 0mm 0mm 0mm 0mm, clip, width=1\linewidth]{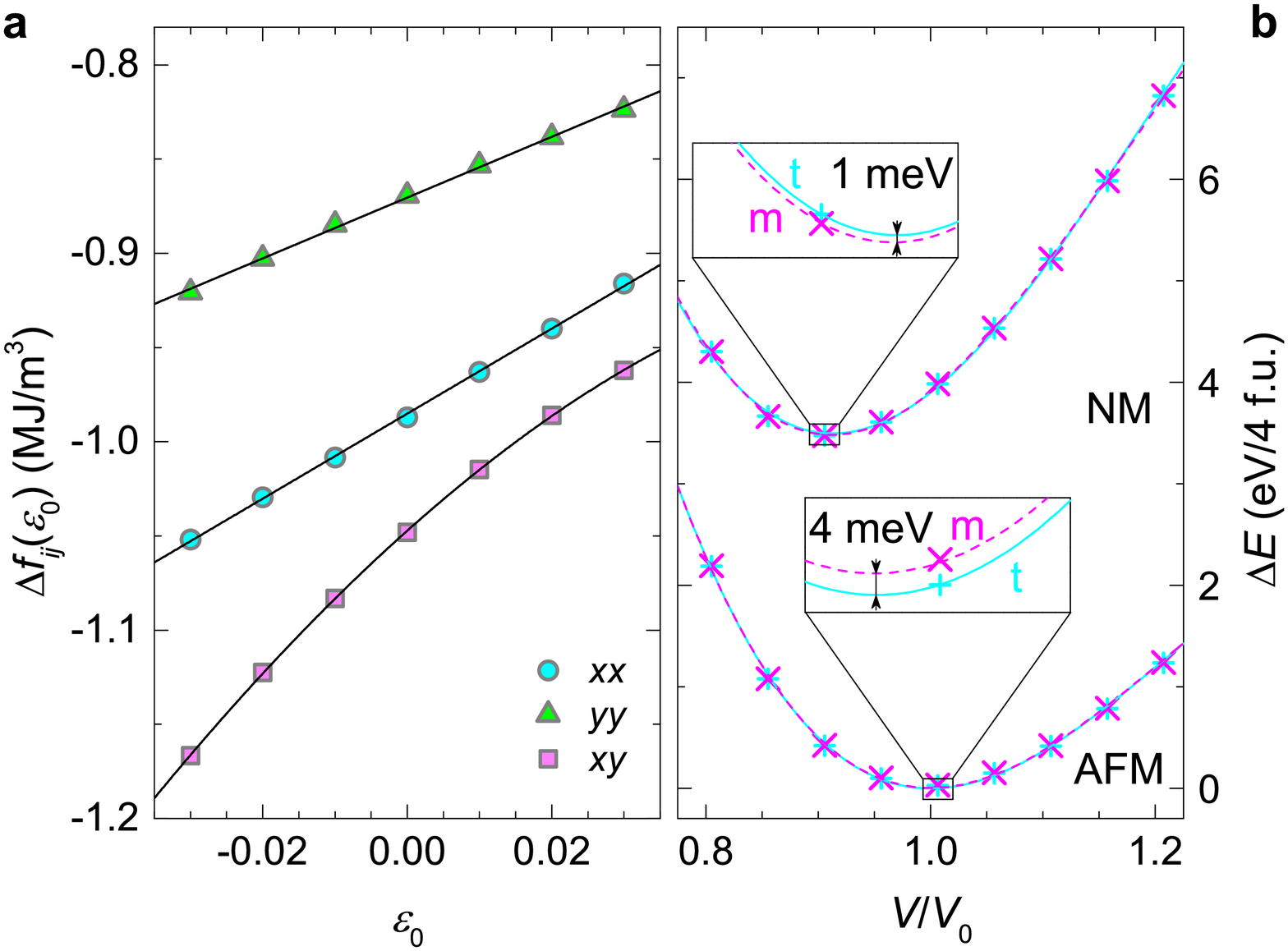}
\caption{{\bf DFT calculations.} {\bf (a)} The calculated difference in the total-energy density for the three different magnetoelastic components. 
The solid lines are linear fits for the $xx$ and $yy$ components, and a quadratic fit for the $xy$ component. 
{\bf (b)} The \textit{ab-initio} calculated total energy of the relaxed monoclinic (m) and triclinic (t) structures of the \Cu~as a function of the volume for the antiferromagnetic (AFM) and non-magnetic (NM) cases. 
The insets zoom at the regions around the local minima of the relaxed structures. The global minimum of the energy is set to zero and the corresponding volume of the triclinic structure $V_0$ is used for volume normalization.}
\label{fig4}
\end{figure}

\begin{figure*}[t]
\includegraphics[trim = 0mm 0mm 0mm 0mm, clip, width=1\linewidth]{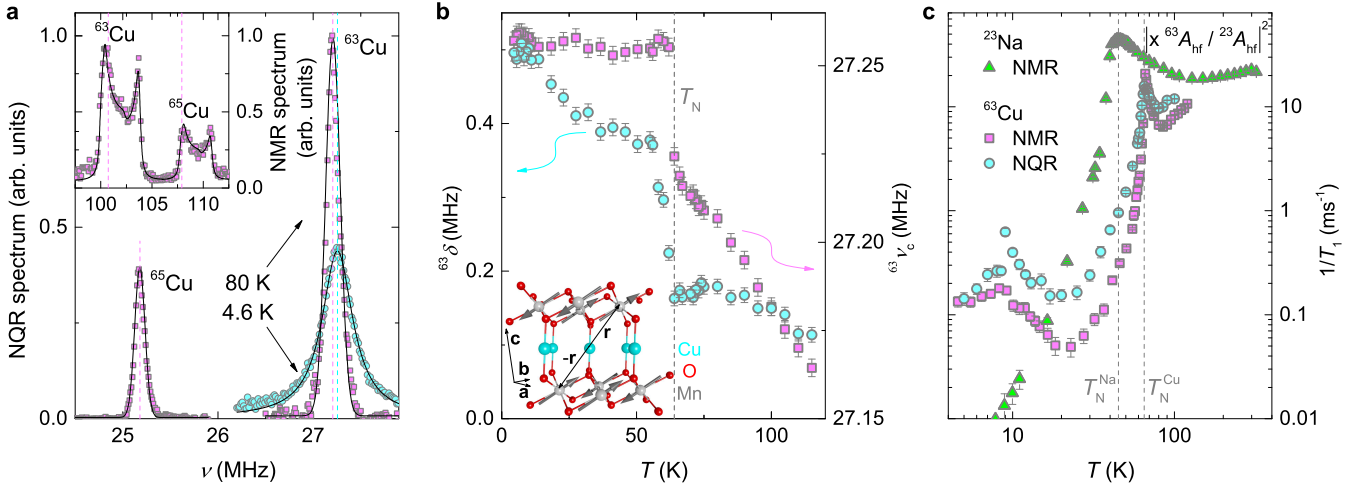}
\caption{{\bf NMR results.} {\bf (a)} 80-K $^{63,65}$Cu NQR and NMR (inset) spectra of \Cu. 
The solid lines represent a simultaneous NMR/NQR fit (see Methods for details), assuming a Gaussian distribution of the NQR frequencies $^{63,65}\nu_{\rm NQR}={^{63,65}\nu_{\rm Q}} \sqrt{1+\eta^2/3}$, that yields the quadrupolar frequency $^{63}\nu_{\rm Q}=27.0(1)$~MHz, the asymmetry parameter $\eta=0.20(5)$, the isotropic hf shift $K_{\rm hf}=1.6(1)\%$ that is much larger than the dipolar shift $K_{\rm d}=0.11\%$, and the individual line widths $^{63}\delta=0.17(1)$~MHz, $^{65}\delta=0.14(1)$~MHz. 
The dashed lines show the center of the NQR lines and the reference NMR frequencies corresponding to a zero magnetic shift.
The $^{63}$Cu NQR spectrum at 4.6~K is added for comparison. 
{\bf (b)} The temperature dependence of the $^{63}$Cu NQR line width $^{63}\delta$ and the line position $^{63}\nu_{\rm c}$. 
The inset highlights the hf paths through the O$^{2-}$ sites that provide the coupling of each Cu nuclei with six surrounding Mn$^{3+}$ magnetic moments (arrows), ordered with the magnetic wave vector \cite{vecchini_magnetoelastic_2010} ${\bf k}=(0,\frac{1}{2},\frac{1}{2})$.
{\bf (c)} Comparison of the temperature-dependent $^{63}$Cu NQR/NMR spin-lattice relaxation rate $1/T_1$ in the \Cu~and $^{23}$Na NMR $1/T_1$ in the \Na~(Ref.~\onlinecite{zorko_frustration-induced_2014}). 
The latter is normalized by the squared ratio of the hf coupling constants. 
The error bars represent the standard deviation of the fit parameters.}
\label{fig1}
\end{figure*} 

\vspace{0.5 cm}
\noindent  {\bf Total-energy change at the $T_{\rm N}$.}
Having established that the magnetic ordering at the $T_{\rm N}$ is predominantly set by the 2D Heisenberg Hamiltonian and the tendency of both systems to remove magnetic degeneracy in the ground state by lattice deformation, the question that arises is what is the microscopic origin of such a complex transformation.
In this respect, the ME coupling has been suggested as being the key factor \cite{giot_magnetoelastic_2007, damay_spin-lattice_2009, vecchini_magnetoelastic_2010} in both \Na~and \Cu.
However, the ME coupling has been shown to be insubstantial in the former case \cite{zorko_frustration-induced_2014}, and thus needs to be evaluated also in the \Cu.
The total-energy change at the $T_{\rm N}$, associated solely with the magnetoelasticity, arises from the coupling terms \cite{du_tremolet_de_lacheisserie_magnetostriction_1993} $b_{ij}\epsilon_{ij}m_i m_j$ 
between the strain-tensor components $\epsilon_{ij}$ and the 
magnetization-direction vector ${\bf m}=\left(m_x,m_y,m_z \right)$. 
The strength of the ME coupling, and consequently the corresponding contribution to the total energy, is proportional to the ME-coupling coefficients $b_{ij}$ ($i,j=x,y,z$).
The coefficients $b_{xx}=2.3$~MJ/m$^3$, $b_{yy}=1.6$~MJ/m$^3$ and $b_{xy}=3.4$~MJ/m$^3$ are determined as linear terms in the calculated dependence of the total-energy-density change $\Delta f_{ij}$ on the strain (see Methods), which is shown in Figure~\ref{fig4}a. 
The ME energy gain is the largest for the shear-strain component $\epsilon_{xy}$, which is associated with the monoclinic-to-triclinic deformation. 
However, for the experimental strain $\epsilon_{xy}=0.0028$ (see Methods) the magnetoelastic energy change amounts to only 2.7~$\mu$eV per triclinic unit cell, which is very similar to the value of 2.5~$\mu$eV found in \Na.

The contribution of the ME coupling to the total energy change at the $T_{\rm N}$ is thus negligible in both compounds. 
Therefore, the complex phase transition at the $T_{\rm N}$ has to reflect changes in the magnetic-exchange and elastic energies \cite{zorko_frustration-induced_2014}.
Our \textit{ab-initio} calculations of both relevant contributions to the total energy in the \Cu~for the non-spin-polarized case, relevant to non-magnetically ordered structures, reveal that the monoclinic structure is energetically lower than the triclinic one, although only by $\sim $1~meV per 4 formula units (f.u.); see Figure~\ref{fig4}b. 
This is in-line with the $C2/m$ crystal symmetry found experimentally at room temperature.
However, once the magnetic order sets in, the total energy of the triclinic structure is lowered below that of the monoclinic structure by about 4~meV per 4 f.u.
This change of $\sim$1.25~meV per Mn$^{3+}$ ion is mainly a consequence of the exchange-energy decrease during the structural phase transition, associated with the removal of the degenerate magnetic states due to the interchain frustration.
The resulting splitting of the interchain exchange constants by \cite{jia_magnetic_2011} $J_{\rm 2a}-J_{\rm 2b}=0.4$~meV releases $S^2(J_{\rm 2a}-J_{\rm 2b})=1.6$~meV of energy per Mn that is slightly larger than the total-energy change at the $T_{\rm N}$, as it is partially spent to compensate for the elastic-energy increase in the triclinic phase.

The calculated total-energy difference below the $T_{\rm N}$ of 1~meV per f.u.~in \Cu~between the two structures that is about 3-times above the calculation error bar (see Methods), is markedly larger than in the \Na, where the calculated difference was below the calculation error bar; $\lesssim$0.5~meV per f.u. \cite{ouyang_first-principles_2010}.
However, in absolute terms this difference is small, even in the \Cu, so that a competition between the near-degenerate monoclinic and triclinic structures is expected for both compounds. 
Experimental local-probe magnetic techniques are then essential for highlighting possible differences between the two systems. 

\vspace{0.5 cm}
\noindent {\bf NMR/NQR insight to the magnetism.}
Information about the magnetic properties of \Cu~on the local scale are revealed in the NQR/NMR experiments via the hyperfine (hf) coupling $A_{\rm hf}$ of the electronic and the $^{63,65}$Cu nuclear magnetic moments. 
Moreover, the quadrupolar splitting in the electric-field gradient (EFG) provides information about the material's structural properties. 
The NQR spectra measured in zero field correspond to a single line for each copper isotope \cite{abragam_principles_1961}, while the powder NMR spectra are structured (see Methods for details).
Our simultaneous fit of the NQR and NMR data at 80~K (Figure~\ref{fig1}a) yields a hf coupling constant $^{63}A_{\rm hf}=2.3(1)$~T/$\mu_B$ that is significantly larger than the coupling constant $^{23}A_{\rm hf}=0.11(1)$~T/$\mu_B$ found \cite{zorko_frustration-induced_2014}  in \Na.
Since $A_{\rm hf}$ scales with the orbital overlap, the charge transfer from the Mn$^{3+}$ ions to the interlayer cations (Cu$^+$ or Na$^+$) is much larger in the \Cu~than in the \Na.
This implies a stronger $J'$ in the former compound and thus is in line with the somewhat better agreement between the experimental and the predicted $T_{\rm N}$, as the renormalization factor $k$ in equation~(\ref{eq5}) is closer to the used value of 0.7.
\begin{figure*}[t]
\includegraphics[trim = 0mm 0mm 0mm 0mm, clip, width=1\linewidth]{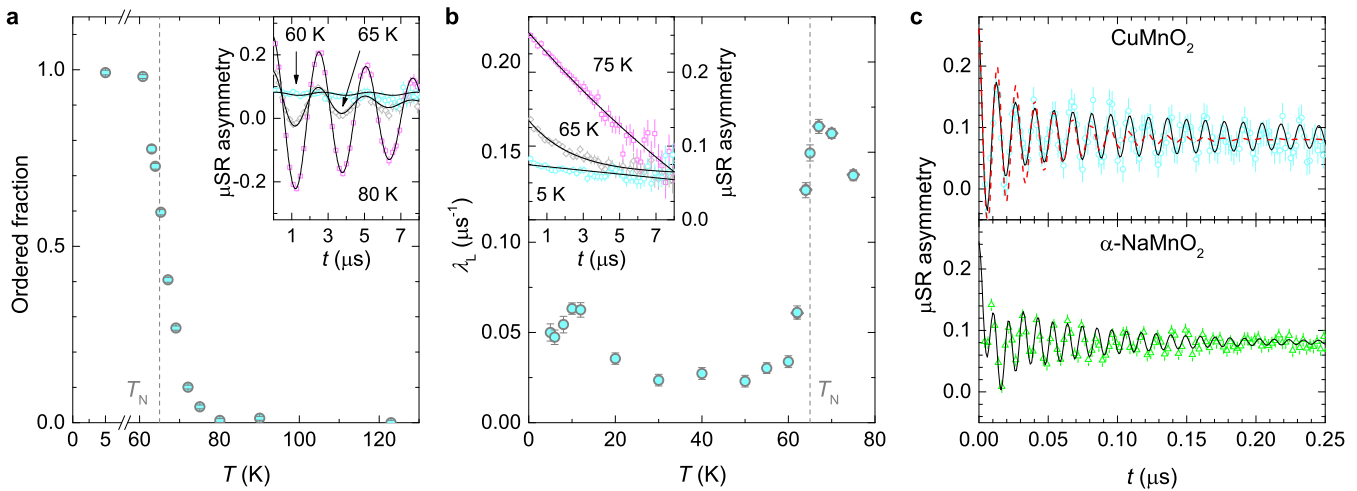}
\caption{{\bf $\mu$SR results.} {\bf (a)} The temperature-dependent magnetically ordered volume fraction $\left[1-A_0(T)/A_0(120\;{\rm K})\right]$ of the \Cu, derived from the wTF $\mu$SR asymmetry $A_{\rm wTF}$ data (inset); solid lines are fits to the model 
$A_{\rm wTF}(t)=A_0(T)\cos \left(\gamma_\mu B_{\rm wTF} t \right){\rm e}^{-\lambda_T t}+C(T)$,
where $\gamma_\mu=2\pi\times 135.5$~MHz/T is the muon gyromagnetic ratio, $B_{\rm wTF}$ is the transverse applied magnetic field and $\lambda_T$ the transverse muon relaxation rate.
The $A_0(T)$ term corresponds to muons experiencing no sizeable static internal magnetic field, while the $C(T)$ term describes those muons that reside at sites with large static fields ($B_\mu \gg B_{\rm wTF}$).
{\bf (b)} The longitudinal muon relaxation rate derived from the ZF muon asymmetry $A_{\rm ZF}$ data (inset); solid lines are fits to the model 
$A_{\rm ZF}(t)=A'_0(T) {\rm e}^{-\lambda_L t}$,
where the initial asymmetry $A'_0(T)$ is temperature dependent to account for the disappearance of the oscillating component below the $T_{\rm N}$.
{\bf (c)} $\mu$SR asymmetry of \Cu~(upper panel) and \Na~(lower panel; adopted from Ref.~\onlinecite{zorko_frustration-induced_2014}) at 5~K. 
The solid and dashed lines represent the corresponding fits to the "two-component" model of equation~(\ref{eq3}) and a model with only one oscillating component, respectively. 
Fitting to the \Cu~data yields $\chi^2=0.71$ and 1.71 for the former and the latter models, respectively. 
The error bars represent the standard deviation of the fit parameters. 
For the muon asymmetry data the latter are set by the square root of the total number of detected positrons.}
\label{fig2}
\end{figure*} 

The width $\delta$ of the NQR spectra at 80~K, amounting to $0.16\text{ MHz}/27.2\text{ MHz} = 0.6\%$ of the line-position value $\nu_c$ already in the paramagnetic phase (Figure~\ref{fig1}b), is rather large.
The line widths $^{63}\delta>{^{65}\delta}$ reveal spectral broadening being in accordance with the quadrupole moments $^{63}Q>{^{65}Q}$ and contradicting the gyromagnetic ratios $^{63}\gamma<{^{65}\gamma}$.
Therefore, sizeable structural distortions of the local environments must be present. 
The temperature dependence of both $\nu_c$ and $\delta$ shows a pronounced sudden increase below the $T_{\rm N}$ (Figure~\ref{fig1}b), clearly marking the phase transition. 
The anomaly in $\nu_c$ at the $T_{\rm N}$ is attributed to the structural transformation of the \Cu~sample, directly affecting the quadrupolar frequency $\nu_{\rm Q}$.
Namely, static internal magnetic fields below the $T_{\rm N}$ cause a symmetric broadening/splitting of the NQR line so that its center of gravity is unaffected \cite{abragam_principles_1961}. 
On the other hand, the pronounced increase of $\delta$ by a factor of $\sim$2 at the $T_{\rm N}$, exceeding the change of $\nu_c$ by several orders of magnitude, can only be magnetic in origin. 

We must emphasize that the existence of the NQR signal below the $T_{\rm N}$ is unexpected.
Namely, in the frame of the homogeneously ordered magnetic phase \cite{vecchini_magnetoelastic_2010} with ${\bf k}=(0,\frac{1}{2}, \frac{1}{2})$ the Cu nuclei would experience extremely large internal magnetic fields.
Although the Cu site is a structural center of inversion, the magnetic order breaks this symmetry, as the spins at $\pm {\bf r}$ from a given Cu site are FM ordered (see the inset in Figure~\ref{fig1}b), in contrast to the \Na, where the order of the two corresponding spins is AFM \cite{zorko_frustration-induced_2014}.
Such a spin configuration in \Cu~yields a large local hf field $B_{\rm hf}={^{63}A_{\rm hf}/3} \times \mu \simeq 2.2$~T ($\mu=3.05 \mu_{B}$ is the size of the ordered \cite{damay_spin-lattice_2009} Mn$^{3+}$ moment). This field leads to extremely broad NQR spectra, $^{63}\delta_{\rm} = {^{63}\gamma}/2\pi\times 2B_{\rm hf} \simeq 50$~MHz, being two orders of magnitude broader than the experimental ones.
Indeed, the NQR signal below the $T_{\rm N}$ corresponds to a minority fraction of all the $^{63}$Cu nuclei, while a majority of the signal is lost at the $T_{\rm N}$ due to the onset of large internal fields. 
Namely, the Boltzman-corrected intensity of the NQR signal at 4.6 K, when further corrected for nuclear relaxation effects, is smaller than the intensity at 80~K by a factor of $\sim17$. 
This reveals that, unexpectedly, about 6\% of all the Cu sites in our sample experience small or no internal magnetic fields and do not correspond to the reported homogeneous magnetic phase. 
We note that the AFM order of the moments positioned symmetrically with respect to the Cu site, or the absence of any order, result in a zero static local magnetic field at the Cu site, and would explain the NQR-observable sites below the $T_{\rm N}$.
Since this minority signal exhibits clear anomalies at the $T_{\rm N}$ (Figure~\ref{fig1}b) it is obviously well coupled to the bulk that undergoes the magnetostructural transition.  
This is confirmed by the temperature dependence of the spin-lattice relaxation rate, 1/$T_1$, that shows a maximum at the $T_{\rm N}$ due to critical spin fluctuations \cite{moriya_nuclear_1956} related to the magnetic instability of the bulk.
However, in contrast to the monotonic decrease found in the Na-based compound below the $T_{\rm N}$, in the \Cu, another clear maximum in both the NMR and NQR  1/$T_1$ is observed at around 10~K.
This reveals an, as yet, unobserved instability that could be either magnetic or structural in its nature. 

\vspace{0.5 cm}
\noindent  {\bf Probing the magnetic disorder with $\mu$SR}.
In order to provide more insight into the magnetic state in the \Cu~that is, according to the unexpected minority NQR signal, apparently not as homogeneous as inferred from previous bulk measurements, we resorted to the $\mu$SR local-probe technique. 
Moreover, this technique reveals details about the low-temperature anomaly in the NMR/NQR relaxation at 10~K.
In contrast to NQR/NMR, which is limited because the intrinsic signal disappears below the $T_{\rm N}$, the $\mu$SR measurements can assess the magnetic properties of the entire \Cu~sample also below the $T_{\rm N}$.
This time a hf/dipolar coupling between the electronic magnetic moments and the muon magnetic moment is utilized after a muon stops in the sample.
The resulting local magnetic field $B_\mu$ at the muon site affects the $\mu$SR asymmetry $A(t)$ that is proportional to the muon polarization precessing in $B_\mu$.  

In \Cu, the weak-transverse-field (wTF) experiment that effectively keeps track of the temperature-dependent ordered part of the sample by measuring the amplitude of the oscillating $\mu$SR signal \cite{yaouanc_muon_2011}, reveals that the fraction of the muons detecting large frozen internal fields starts growing already below 80~K (Figure~\ref{fig2}a); i.e., far above the $T_{\rm N}$, which can be attributed to developing short-range spin correlations \cite{yaouanc_muon_2011}.
Below the $T_{\rm N}$, the whole sample becomes magnetically ordered within only a few kelvins, leaving no room for a non-frozen fraction above the experimental error bar of a few percent.   
Similar information is obtained from the zero-field (ZF) $\mu$SR, where the initial asymmetry strongly decreases around the $T_{\rm N}$ and at low temperatures reaches 1/3 of its high-temperature value (inset in Figure~\ref{fig2}b). 
Such a reduction is characteristic of the establishment of strong static internal fields in powder samples.
Statistically, in 1/3 of all cases the muon magnetic moment is aligned parallel to the $B_\mu$ and therefore exhibits no precession, while rapid oscillations of the asymmetry in other cases diminish the $\mu$SR signal on a coarse time scale. 

A detailed look at the ZF relaxation curves below $T_{\rm N}$ (Figure~\ref{fig2}c) also allows for the detection of the quickly-oscillating component. 
Similar to the \Na~case \cite{zorko_frustration-induced_2014}, the two-component model
\begin{equation}
A_{\rm ZF}(t) = \sum_{j=1}^{2}  {f_j\left[\frac{1}{3}{\rm e}^{-\lambda_L t}+\frac{2}{3}{\rm cos}(\gamma_\mu B_{\mu,j}t){\rm e}^{-\lambda_{T,j} t}\right]}
\label{eq3}
\end{equation}
fits well with the experimental data.     
Here, $f_j$ denotes temperature-independent probabilities that the muons stop at either of the two magnetically non-equivalent stopping sites $j$.
The preferential site is occupied in 70(5)\% cases.
The internal field at this site is only slightly higher than at the second site (0.59 and 0.54~T at the first and the second site, respectively, at 5 K); however, a fit with only a single oscillating component (the dashed line in Figure~\ref{fig2}c) results in a much poorer agreement with the data. 
The damping rate of the oscillations $\lambda_{T,j}$ that is due to the finite width of the local-field distributions \cite{zorko_frustration-induced_2014} in \Cu~is reduced by a factor of $\sim$3 when compared to the \Na~(Figure~\ref{fig2}c), indicating more homogeneous magnetism.  

In the ZF experiment, the magnetic phase transition at the $T_{\rm N}$ is expressed as a maximum of the longitudinal muon-relaxation rate $\lambda_{\rm L}$, like in the NQR/NMR relaxation experiments. 
Moreover, the second maximum observed in the NQR/NMR experiments at 10~K is also found in the $\mu$SR. 
Since the ZF $\mu$SR signal corresponds to the total volume of the sample and the muons are only sensitive to magnetism, this reveals that the low-temperature anomaly is of magnetic origin and is intrinsic to the \Cu~system.

\section*{Discussion} 
The FTLM and DFT numerical calculations provide a solid basis for addressing the experimentally observed similarities and differences between the \Cu~and the \Na. 
Considering the latter calculations in the magnetically ordered state, the triclinic phase is energetically preferred in both compounds.
With increasing temperature, the staggered susceptibility decreases and this leads to a finite $T_{\rm N}$.
Above the $T_{\rm N}$ the exchange-energy gain associated with the magnetically ordered state disappears, which in turn leads to a structural transformation to the monoclinic phase that is energetically preferred in the non-magnetic state. 
The isotropic Heisenberg Hamiltonian of the spatially anisotropic triangular lattice is dominantly responsible for elevating the $T_{\rm N}$ in the \Cu~with respect to the \Na, while the magnetoelastic and the interlayer couplings play a less important role.

On the other hand, our local-probe experiments on the \Cu revealed some subtle, yet profound, features that should be carefully considered in the attempt to understand the presence/absence of nano-scale phase separation in the spatially anisotropic triangular lattice.
Both, the NQR/NMR and the $\mu$SR investigations demonstrated that \Cu~undergoes a magnetostructural phase transition at $T_{\rm N}=65$~K almost completely. 
The minority NQR component ($\sim$6\%) that remains present below the $T_{\rm N}$ can be explained by regions where the interlayer magnetic ordering is FM instead of being AFM, as the latter causes the disappearance of the NQR signal due to large local fields.
The NQR signal exhibits a magnetic anomaly around 10~K, which is expressed by the increased relaxation rates of the NQR/NMR as well as the $\mu$SR.  
Since $\mu$SR, on the other hand, detects a bulk magnetic signal,
the small NQR component is apparently coupled to the bulk magnetic phase.
This is further confirmed by the line position and the width of the NQR spectra, changing considerably at the $T_{\rm N}$. 
The coupling with the bulk phase can then be regarded in the context of the nano-scale phase inhomogeneity.  
A comparison of the ZF $\mu$SR asymmetry curves of the \Cu~and the \Na~is quite informative in this respect.
The notably reduced damping of the oscillations in the ordered phase of the former compound provides evidence of much narrower field distributions, and hence less disorder. 
This conclusion is also in line with the number of the interlayer cation (Cu$^+$ and Na$^+$) sites experiencing internal fields that do not comply with the symmetry of the bulk magnetic order, which in the \Cu~is decreased to 6\%, from the 30\% found \cite{zorko_frustration-induced_2014} in the \Na.

The magnetostructurally inhomogeneous ground state of the \Na~on the nano-scale has previously been attributed to the combined effects of geometrical frustration and near-degenerate monoclinic and triclinic structural phases \cite{zorko_frustration-induced_2014}.
We believe that the key factor controlling such an inhomogeneity is the difference in the total energy of the two competing phases in the magnetically ordered state. 
This difference is notably larger in the \Cu~(1~meV per f.u.) than in the \Na, where it is below the computational error bar (<0.5~meV per f.u. \cite{ouyang_first-principles_2010}). 
In the latter compound, an infinitesimal quenched disorder, locally favouring one phase over the other, can then be held responsible for triggering the phase separation. 
Similar effects are suppressed in the stoichiometric \Cu, but would become enhanced for larger deviations from perfect system uniformity.
Indeed, enhanced strain, acting as a precursor of the monoclinic-to-triclinic structural phase transition, has been observed \cite{zorko_frustration-induced_2014, giot_magnetoelastic_2007} in the high-temperature monoclinic phase in stoichiometric \Na, while in the \Cu~a Cu-Mn off-stoichiometry is required to produce such a strain \cite{poienar_substitution_2011}.
Moreover, the diffuse magnetic scattering characteristic of 2D correlated regions that coexist with sharp magnetic Bragg peaks (one of the signatures of the inhomogeneity \cite{zorko_frustration-induced_2014} found in the \Na) is also found \cite{damay_spin-lattice_2009,garlea_tuning_2011, terada_magnetic_2011} in the \Cu. 
However, in contrast to the \Na, where it persists to low temperatures, in stoichiometric \Cu~it gradually gives way to the 3D ordered phase below the $T_{\rm N}$. 
Interestingly though, in off-stoichiometric samples \cite{garlea_tuning_2011} the volume fraction of the 2D-correlated phase shows no decrease below the $T_{\rm N}$, implying that the 2D-ordered regions keep competing with the 3D order at low temperatures. 
The total-energy difference of the competing phases below the magnetostructural transition, reflecting the interplay of the magnetic-exchange and the elastic energies, then seems to determine the amount of disorder required to stabilize the inhomogeneous ground state on a geometrically frustrated triangular lattice.
Systems with near-degenerate competing phases can be locally perturbed more easily. 
Such an inhomogeneity may, therefore, be a more general feature of geometrically frustrated magnets.

\section*{Methods}
\noindent {\bf Finite-temperature Lanczos method simulations.}
Calculations of the spin susceptibility for the $S=2$ Heisenberg model on the anisotropic triangular lattice (equation~(\ref{eq4})), were performed with the finite-temperature Lanczos method (FTLM) \cite{jaklic00,prelovsek13} and were used to determine the leading exchange couplings $J_1$ and $J_2$ in the \Cu. Within the FTLM finite-size clusters are diagonalized in a similar manner as for the standard exact diagonalization Lanczos method (at $T=0$) and additional random vector averaging over the $R$ vectors is employed to determine the properties at $T>0$. 
Typically, $R\sim 10$ suffices for the largest systems and the lowest $T$, while smaller systems require a larger $R$.  
The limitations of the method are mainly set by finite-size effects, which are the largest at low $T$ and determine the lowest reachable $T$($\sim J_1$). 
In order to reduce the finite-size effects we used periodic boundary conditions, adjusted cluster shapes, the largest reachable cluster sizes (up to $N=14$ sites), and additional approximations for the values in the thermodynamic limit.

\begin{figure}[t]
\includegraphics[trim = 0mm 0mm 0mm 0mm, clip, width=1\linewidth]{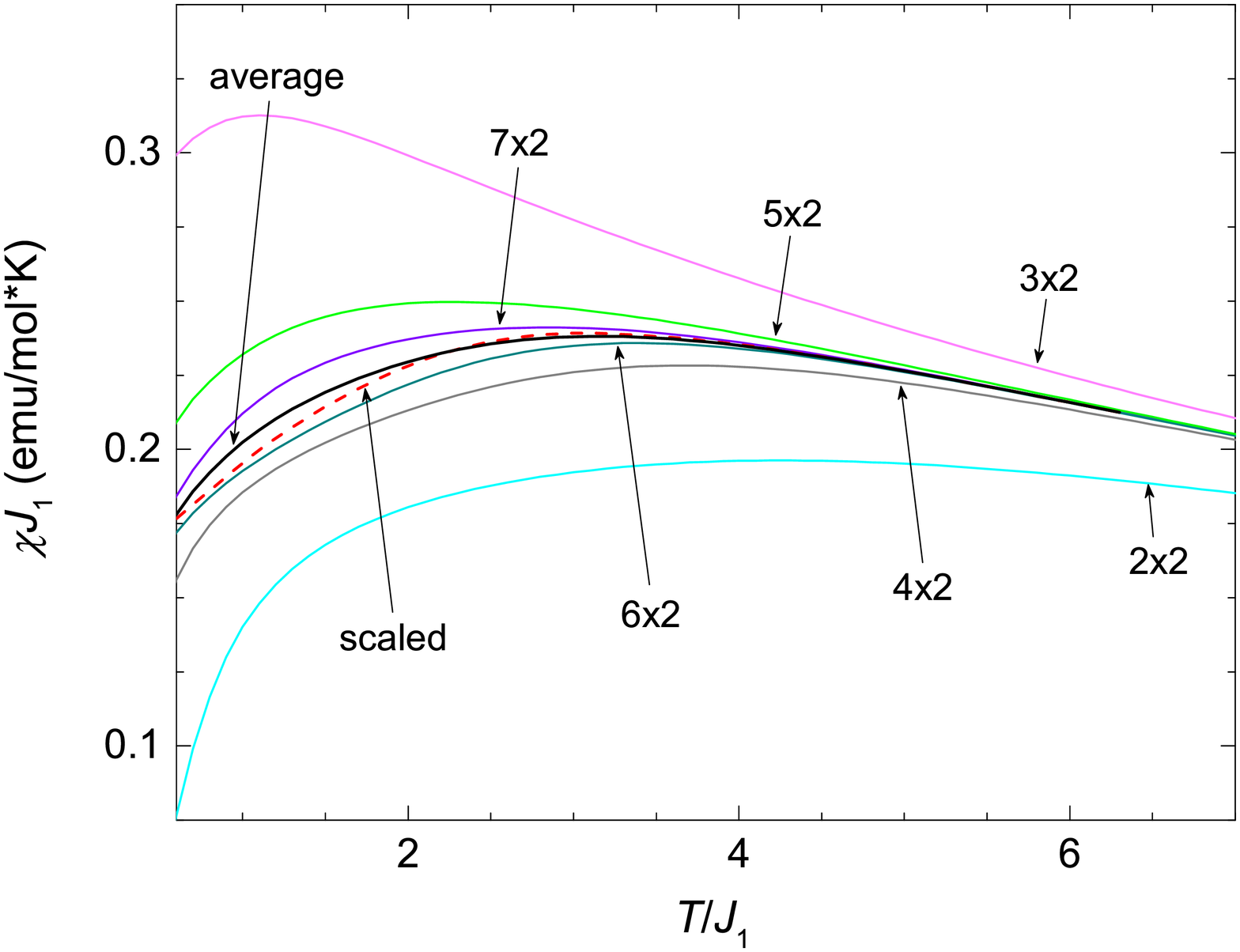}
\caption{{\bf The finite-size effects in the FTLM calculation.}  The calculated susceptibility on various $m \times n$ clusters for the optimal parameters $J_1=53$~K and $J_2/J_1=0.27$. The curve averaging the $6\times2$ and $7\times2$ data and the scaled curve are also shown.}
\label{figA}
\end{figure}

The temperature-dependent magnetic susceptibility $\chi(T)$ of the model given by equation~(\ref{eq4}) was calculated previously in Ref.~\onlinecite{zorko_magnetic_2008} for \Na. 
It was shown that in the regime of interest, elongated spin clusters are the most appropriate. 
In particular, if a cluster has $m$ independent spins in the $J_1$ direction and $n$ spins in the $J_2$ directions, it was realized that due to $J_2<J_1$ and two competing $J_2$ bonds, $\chi$ does not depend on $n$ for $n>2$ (see Fig.~2 in Ref.~\onlinecite{zorko_magnetic_2008}). 
This fact allows us to reduce the finite-size effects by using a larger $m$.
We note that due to the alternating behaviour of $\chi$ with $m$ (see Figure~\ref{figA}), which originates in periodic boundary conditions and antiferromagnetic spin-spin correlations, the average susceptibility curve from the two largest-size clusters ($m=6$ and $7$) is a better approximation than the $m=7$ curve.

\vspace{0.5 cm}
\noindent {\bf Scaling-like approximation for the susceptibility.}
The results for several different sizes of finite clusters and their systematics, shown in Figure~\ref{figA}, allow a scaling-like analysis to obtain a better approximation of $\chi$ in the thermodynamic limit. 
Typical scaling analyses use scaling functions of the form $\chi(N)=a+b/N$ and additional higher terms when needed; e.g., $c/N^2$. 
Since our calculations are limited to rather small maximum system sizes by $S=2$, we also use the results from small systems (starting with $m=2$).
Consequently, such scaling functions are not appropriate. 
In particular, at higher $T$ (see, for example, $T>5J_1$ in Figure~\ref{figA}) $\chi$ has already converged with $N$ for systems with $m\ge 5$, while for $m<5$ notable finite-size effects are seen. 
Therefore, the scaling function should be close to a constant for $1/N$ smaller than some value, while at larger $1/N$, the scaling function should allow for a stronger $N$ dependence. 
For these reasons we use a generalized scaling function of the form $\chi(N)=a+b\left[\exp(c/N)-1\right]$, which corresponds to typically used functions in the limit of small $1/N$. 
In order to also capture the alternating component of $\chi$ with $m$ (Figure~\ref{figA}) we add, in a similar fashion, the term $b_1\left[\exp(c_1/N)-1\right](-1)^{N/2}$.
Such a scaling function also gives a correct (converged with $N$) result for high $T$, while typical scaling functions fail in this respect.
We have performed such a scaling for each $T$ separately. 
However, since we are limited to small systems with notable finite-size effects at low $T$ ($\lesssim J_1$) and since the scaling function has many parameters, the result of such an analysis should not be taken as a strict thermodynamic limit. 
Rather, it should be regarded as a next approximation of it, compared to the result from simpler averaging of the two largest-cluster curves.  

\vspace{0.5 cm}
\noindent {\bf Density-functional theory calculations.}
The calculations of the total energies and the magneto-elastic (ME) coupling coefficients were performed within the framework of the density-functional theory (DFT) and the generalized-gradient approximation (GGA) \cite{Perdew:1996} for the exchange-correlation 
contribution by applying the Quantum Espresso code \cite{Giannozzi:2009}.
The electron-ion interactions were described by the Vanderbilt ultrasoft potentials \cite{Vanderbilt:1990} including the spin-orbit coupling for the Mn atoms. The plane-wave cut-off parameters were set to 585 eV and 4678 eV for the expansion of the wave functions and the potential, respectively.
In order to take into account the proper antiferromagnetic ordering, 
the $1\times 2\times 2$ supercells of the monoclinic and the triclinic structures were used. 
The calculations of the total energies as a function of the unit-cell volume for the different types of magnetic ordering were carried out by using $4\times 8\times 2$ reciprocal vectors in the full Brillouin zone (BZ) for the Methfessel-Paxton sampling  \cite{Methfessel:1989} integration.
The criterion for the self consistency was the total-energy difference 
between two subsequent iterations being less than $10^{-8}\>{\rm Ry}$.
The monoclinic phase was further optimized by minimizing the total energy and the inter-atomic forces with respect to the lattice parameters and the atomic positions. 
The resulting structure served as the zero-strain reference for the calculations of the ME coefficients that are based on the evaluation of the total-energy differences of the order of $<10^{-4}\>{\rm Ry}$, which is also the accuracy for the determination of the total-energy differences between the monoclinic and triclinic phases in Figure~\ref{fig4}b, calculated per 4 f.u. 
The tests yielded $8\times 16\times 4$ reciprocal vectors in the full BZ to be enough for well-converged results.

\vspace{0.5 cm}
\noindent {\bf Determination of the magnetoelastic coupling.}
The magnetoelastic coupling constants $b_{ij}$ are calculated from the associated magnetoelastic energy density.
This contains the products $b_{ij}\epsilon_{ij}m_im_j$ of the strain-tensor components $\epsilon_{ij}$ and the components $m_i$ of the normalized magnetization. 
The form of the magnetoelastic energy density is determined by the symmetry of a particular system \cite{du_tremolet_de_lacheisserie_magnetostriction_1993}.
The magnetism of the \Cu~and the \Na~is essentially two-dimensional; therefore, only the terms with the lateral strain-tensor components are important. For the monoclinic symmetry, these include
\begin{eqnarray}
b_{xx} \epsilon_{xx}m_x^2, \nonumber \\
b_{xy} \epsilon_{xy}m_x m_y, \nonumber \\
b_{yy} \epsilon_{yy}m_y^2.
\label{EqB1}
\end{eqnarray}
Individual magnetoelastic terms are then determined by specifically choosing strain components and magnetization directions and calculating the total-energy density $f(\epsilon_{xx},\epsilon_{yy},\epsilon_{xy},m_x,m_y,m_z)$, from
\begin{eqnarray}
\Delta f_{xx}=f(\epsilon_0,0,0,1,0,0)-f(\epsilon_0,0,0,0,0,1), \nonumber \\
\Delta f_{yy}=f(0,\epsilon_0,0,0,1,0)-f(0,\epsilon_0,0,0,0,1), \nonumber \\
\Delta f_{xy}=f(0,0,\epsilon_0,1,1,0)-f(0,0,\epsilon_0,0,0,1).
\label{EqB2}
\end{eqnarray}
The above total-energy differences are calculated {\it ab-initio} as a function of $\epsilon_0$ for relaxed crystal structures. 
In \Cu, $\Delta f_{xx}$ and $\Delta f_{yy}$ change linearly with increasing strain at least up to $\epsilon_0 = 0.03$, while an additional quadratic term is observed in $\Delta f_{xy}$ (Figure~\ref{fig4}a). 
The experimental strain value $\epsilon_{xy}=0.0028$ that is obtained by calculating the relative shift of the Mn$^{2+}$ ions in the triclinic structure, when compared to the monoclinic structure (based on high-resolution synchrotron XRD data analysis \cite{zorko_frustration-induced_2014}), is an order of magnitude lower.
Therefore, the linear term is dominant for all three contributions and allows the extraction of the three magnetoelastic constants $b_{xx}=2.3$~MJ/m$^3$, $b_{yy}=1.6$~MJ/m$^3$ and $b_{xy}=3.4$~MJ/m$^3$. 

\vspace{0.5 cm}
\noindent {\bf Nuclear magnetic/quadrupolar resonance.}
$^{63,65}$Cu ($I=3/2$) NMR/NQR measurements were performed on a high-quality powder sample with the same phase purity and stoichiometry as in the  study presented in Ref.~\onlinecite{vecchini_magnetoelastic_2010}. 
The NMR/NQR spectra and the spin-lattice relaxation were measured between 4.6~K and 120~K in a magnetic field of 8.9~T (NMR) and in zero magnetic field (NQR) on a custom-built spectrometer.
Frequency sweeping and a solid-echo pulse sequence were used for recording the spectra, while a saturation recovery method was used for measuring the spin-lattice relaxation.
Typical $\pi/2$-pulse lengths were 3.5~$\mu$s and 6~$\mu$s in the NMR and NQR experiments, respectively. 
The reference NMR Larmor frequencies of $^{63}\nu_0=100.728$~MHz and $^{65}\nu_0=107.908$~MHz were determined with a 0.1M NaCl-solution reference by taking into account the gyromagnetic ratios $^{23}\gamma=2\pi\times 11.261$~MHz/T, $^{63}\gamma=2\pi\times 11.295$~MHz/T and $^{65}\gamma=2\pi\times 12.089$~MHz/T.

The NQR spectrum of each isotope is particularly simple, as it is given by a single line \cite{abragam_principles_1961} at 
$^{63,65}\nu_{\rm NQR}={^{63,65}\nu_{\rm Q}} \sqrt{1+\eta^2/3}$, with the ratio of the quadrupolar frequencies $^{63}\nu_{\rm Q}/^{65}\nu_{\rm Q}=1.08$ fixed by the corresponding quadrupolar moments and the EFG tensor $\underline{\rm V}_{ij}$ asymmetry parameter being $\eta=(V_{xx}-V_{yy})/V_{zz}$. 
The NMR spectrum is more complicated, because the applied magnetic field $B_0$ breaks the symmetry in the spin space.
The central-transition ($-1/2\leftrightarrow 1/2$) powder NMR line adopts a characteristic structure because of the angular-dependent NMR shift $K$ from the reference frequencies $\nu_0={\gamma}B_0/2\pi$, $^{63,65}K=\frac{\nu- {^{63,65}\nu_0}}{^{63,65}\nu_0}=K_{\rm hf}+K_{\rm d}+ {^{63,65}K_{\rm Q}}$. 
In analogy \cite{zorko_frustration-induced_2014} to the \Na, we take the hf shift $K_{\rm hf}=\hbar A_{\rm hf} \mu/\nu_0$ ($\hbar$ is the reduced Planck constant) to be isotropic, while the dipolar contribution $K_{\rm d}$ and the quadrupolar \cite{abragam_principles_1961} shift $^{63,65}K_{\rm Q}$ can be accurately calculated.
The former has a uniaxial symmetry and is calculated \cite{zorko_frustration-induced_2014} ($K_{\rm d}=0.11\%$ is the dominant eigenvalue) by taking into consideration all the Mn$^{3+}$ paramagnetic spins around a given Cu site within a sphere large enough to ensure convergence.

A homogeneous life-time broadening of the NQR spectra is negligible. 
The spin-spin relaxation time $^{63}T_2=46$~$\mu$s at 80~K yields $^{63}\delta_h=6.9$~kHz, which is much smaller than the spectral width.
The spin-lattice relaxation is of magnetic origin. We find the isotopic effect $^{65}T_1/^{63}T_1$=0.86 that is in accordance with magnetic relaxation dictating $1/T_1\propto \gamma^2 A_{\rm hf}^2$.

\vspace{0.5 cm}
\noindent {\bf Muon spin relaxation.}
The $\mu$SR investigation was carried out on the General Purpose Surface muon (GPS) instrument at the Paul Scherrer Institute, Villigen, using the same powder sample as in the NMR/NQR experiments.
Zero-field (ZF) and weak-transverse-field (wTF) measurements in a 3~mT magnetic field were performed in the temperature range between 5 and 120 K.
The veto mode was utilized to minimize the background signal.
The ZF $\mu$SR measurements below the $T_{\rm N}$ revealed that each muon stops at one of the two possible non-equivalent stopping sites, like was observed \cite{zorko_frustration-induced_2014} in \Na. 

%

\section*{Acknowledgements}
We acknowledge fruitful discussions with S.~El Shawish. 
This work has been supported in part by the Slovenian Research Agency Program P1-0125.
A.L. acknowledges financial support in the framework of the KRIPIS action, PROENYL research project No.~MIS-448305 (2013SE01380034) that was funded by the General Secretariat for Research and Technology, Ministry of Education, Greece and the European Regional Development Fund (Sectoral Operational Programme: Competitiveness and Entrepreneurship, NSRF 2007-2013)/ European Commission.
The $\mu$SR results are based on experiments performed at the Swiss Muon Source (S$\mu$S), Paul Scherrer Institute, Villigen, Switzerland.

\section*{Author contributions}
A.Z., D.A. and A.L. designed and supervised the project. 
The FTLM simulations were performed by J.K., while the {\it ab-initio} calculations were carried out by M.K. 
The samples were synthesized and characterized by O.A. 
The $\mu$SR experiments were performed by A.Z. and H.L.
The NMR/NQR experiments were conducted and analysed by A.Z., who also wrote the paper.
All authors contributed to the interpretation of the data, discussed the results and reviewed the manuscript.

\section*{Additional information}
\noindent {\bf Competing financial interests.} 
The authors declare no competing financial interests.

\end{document}